\documentclass{PoS}
\usepackage{wrapfig}
\usepackage{caption}
\usepackage{graphicx}
\usepackage{subcaption}
\usepackage{enumitem}
\usepackage{amsmath}
\title{Systematic Uncertainties and Cross-Checks for the NOvA Joint $\nu_\mu$+$\nu_e$ Analysis}

\ShortTitle{Systematics and cross-checks}

\author{Reddy Pratap Gandrajula\thanks{Supported by DOE/Fermilab.}\\
        Michigan State University\\
        E-mail: \email{gandraju@msu.edu}}

\author{Micah Groh\\
        Indiana University\\
        E-mail: \email{mcgroh@iu.edu}}

\author{For the NOvA Collaboration}

\abstract{The physics goals of NOvA are the constraints of neutrino oscillation parameters such as the octant of $\theta_{23}$, $\delta_{\rm{CP}}$, and the neutrino mass hierarchy via a joint fit to $\nu_{\mu}$ and $\nu_{e}$ oscillation spectrum. We do this by propagating $\nu_{\mu}$ from the world's most intense neutrino beam at Fermilab, over a baseline of 810 km to northern Minnesota, USA, and measure the $\nu_{\mu}$ to $\nu_{e}$ oscillation probability. NOvA announced its latest oscillation results, based on 8.85$\times10^{20}$ (6.9$\times10^{20}$) protons on target neutrino (antineutrino) data. Preliminary results for the allowed values of oscillation parameters are: $\Delta m^2_{32}=2.51^{+0.12}_{-0.08}\times 10^{-3} \mathrm{eV}^2$, $\mathrm{sin}^2\theta_{23} = 0.58 \pm 0.03$ (upper octant), and $\delta_{\rm{CP}}=0.17\pi$ with preference to the normal hierarchy. Reliable constraints on these oscillation parameters require a rigorous treatment of systematic uncertainties and thorough cross-checks. In this paper, we present an overview of the treatment of systematic uncertainties as well as cross-checks using muon removed simulations and cosmic muon bremsstrahlung showers.}

\FullConference{Proceedings of the Neutrino 2018, the 28th International Conference on Neutrino Physics and Astrophysics\\
                4--9 June, 2018\\
                Heidelberg, Germany}

\begin{document}
\section{Introduction}

The NuMI\footnote{Neutrinos at the Main Injector} Off-Axis $\nu_{e}$ Appearance experiment (NOvA) \cite{nova,nova2} is the flagship long-baseline neutrino experiment in the United States, designed to study the properties of neutrino oscillations. NOvA consists of two functionally equivalent detectors each located 14.6 mrad off the central axis of the Fermilab NuMI neutrino beam, the world's most intense neutrino beam. The Near Detector is located 1 km downstream from the neutrino production source, and the Far Detector is located 810 km away in Ash River, Minnesota. This long baseline, combined with the ability of the NuMI facility to switch between neutrino and anti-neutrino enhanced beams, allows NOvA to make precision measurements of neutrino mixing angles, constrain the neutrino mass hierarchy, and begin searching for CP violating effects in the lepton sector. We study 4 oscillation channels $\overset{\scriptscriptstyle(-)}{\nu}_\mu \rightarrow 
\overset{\scriptscriptstyle(-)}{\nu}_\mu$ and $\overset{\scriptscriptstyle(-)}{\nu}_\mu \rightarrow 
\overset{\scriptscriptstyle(-)}{\nu}_e$. NOvA released its latest oscillation results from a full detector equivalent exposures of 8.85$\times10^{20}$ protons on target neutrino beam and  6.9$\times10^{20}$ protons on target antineutrino beam~\cite{Neutrino2018,mendez_diana_patricia,back_ashley} collected between February 2014 and April 2018. This paper presents the assessment of systematic uncertainties and cross-check studies done in making these precise measurements. It is organized as follows. In Sec.~\ref{sec:Detectors}, we give brief descriptions of NOvA detectors. Sec.~\ref{sec:oscillations} and Sec.~\ref{sec:systematics} describe the neutrino oscillation measurements and associated systematics. The data-driven cross-checks using muon-removed electrons and muon-removed bremsstrahlung showers are explained in Sec.~\ref{sec:mre} and in Sec.~\ref{sec:mrbrem}. 



\section{The Detectors}
\label{sec:Detectors}

The NOvA detectors were designed for electron identification. The two detectors are functionally equivalent, fine-grained, low-Z (0.18 radiation lengths per layer), liquid scintillator calorimeters made of 65\% active material.
The Far Detector (FD) is 14~ktons and sits on the surface in Minnesota. The Near Detector (ND) is 290~tons placed 300~ft underground at Fermilab. These two detectors consist of layered reflective polyvinyl chloride (PVC) cells,  filled with liquid scintillator arranged in alternating horizontal and vertical planes for 3D reconstruction, to form a tracking sampling calorimeter.   When a charged particle passes through the liquid scintillator, which is comprised primarily of mineral oil solvent with a 5\% pseudocumene admixture and PPO and bis-MSB as secondary fluors\cite{oil}, it produces scintillation light. The scintillation light is picked up by a 0.7~mm wavelength shifting fiber within every cell (each cell is read out individually) which is coupled to a 32 pixel avalanche photo-diode (APD) where the light is collected and amplified.\\

\begin{figure}[!htbp]
\begin{center}
\includegraphics[width=0.99\textwidth]{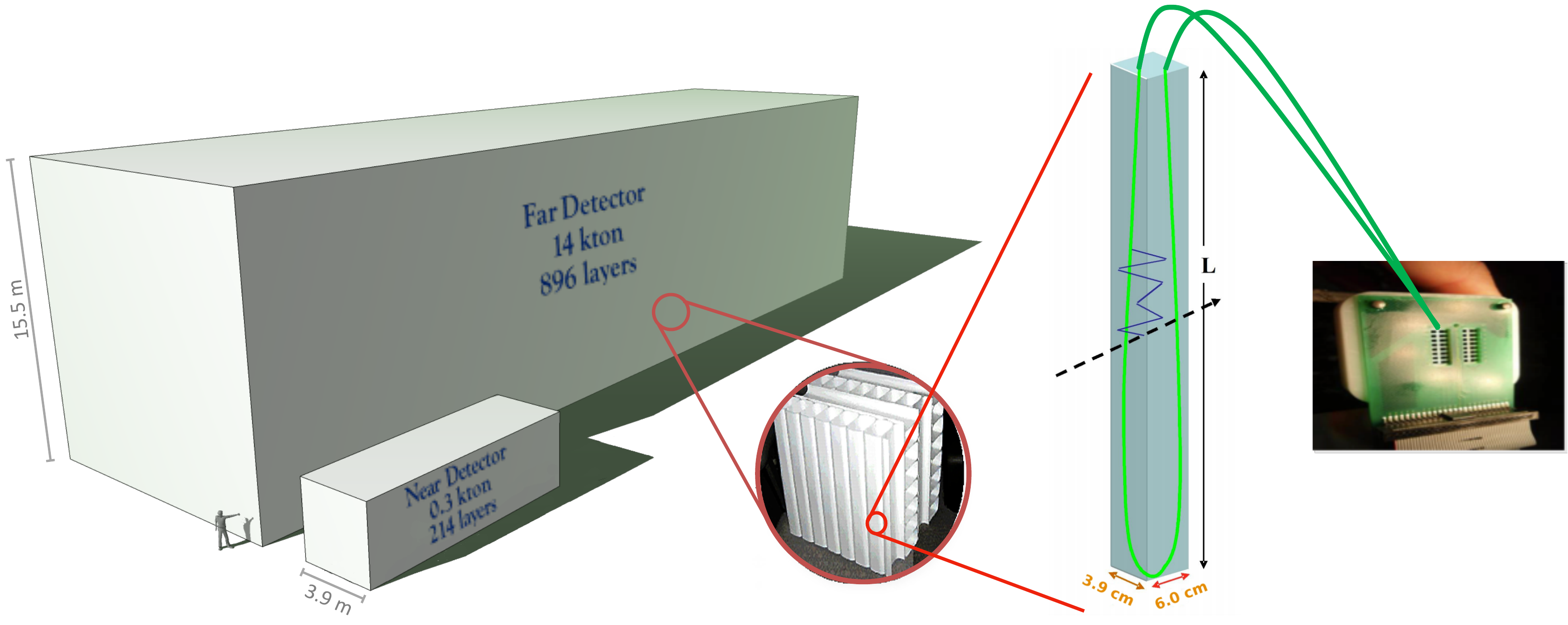}
\caption{The relative sizes of the NOvA Far and Near detectors. The structure of the NOvA detector layers and a single PVC cell coupled to an APD via a wavelength shifting fiber are also shown.}
\label{fig:novadetectors}
\end{center}
\end{figure}

The FD cells are 3.9~cm $\times$ 6.6~cm in cross section, with the 6.6 cm dimension along the beam direction, and 15.5~m long. The ND cells are identical to those of the FD but shorter in length, 3.9~m. In total, there are 344,054 cells in the FD and 21,192 cells in the ND. To improve muon containment, the downstream end of the ND has a "muon catcher" composed of a stack of sets of planes in which a pair of one vertically-oriented and one horizontally-oriented scintillator plane is interleaved with one 10 cm thick plane of steel. The relative sizes of the detectors are shown diagrammatically in Fig.~\ref{fig:novadetectors}. Both detectors are 14.6~mrad off-axis of the NuMI beam. This results in a neutrino flux with a narrow band energy spectrum centered around 2 GeV. Such a spectrum emphasizes $\nu_\mu\rightarrow\nu_e$ oscillations for the NOvA baseline and reduces backgrounds from higher energy neutral current interaction events.

\begin{figure}[!htbp]
\begin{center}
\includegraphics[width=0.99\textwidth]{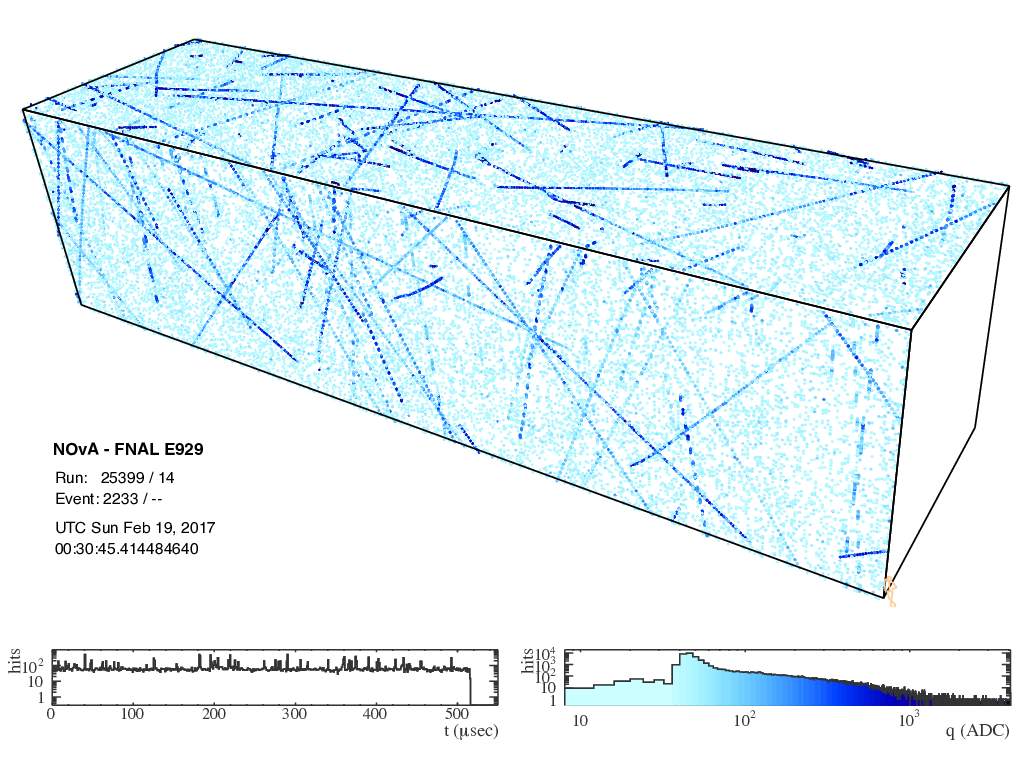}
\caption{5 ms data readout from the NOvA Far Detector.}
\label{fig:5msreadout}
\end{center}
\end{figure}

\section{Neutrinos in NOvA}\label{sec:oscillations}

The NuMI beam at Fermilab creates a "spill" of neutrinos every 1.3 seconds. Each spill lasts for only 10~$\mu$s. NOvA data is recorded in 550~$\mu$s intervals centered on the beam spill, in which hundreds of particle tracks can be seen as shown in Fig.~\ref{fig:5msreadout}.\\

\begin{figure}[!htbp]
\begin{center}
\includegraphics[width=0.49\textwidth]{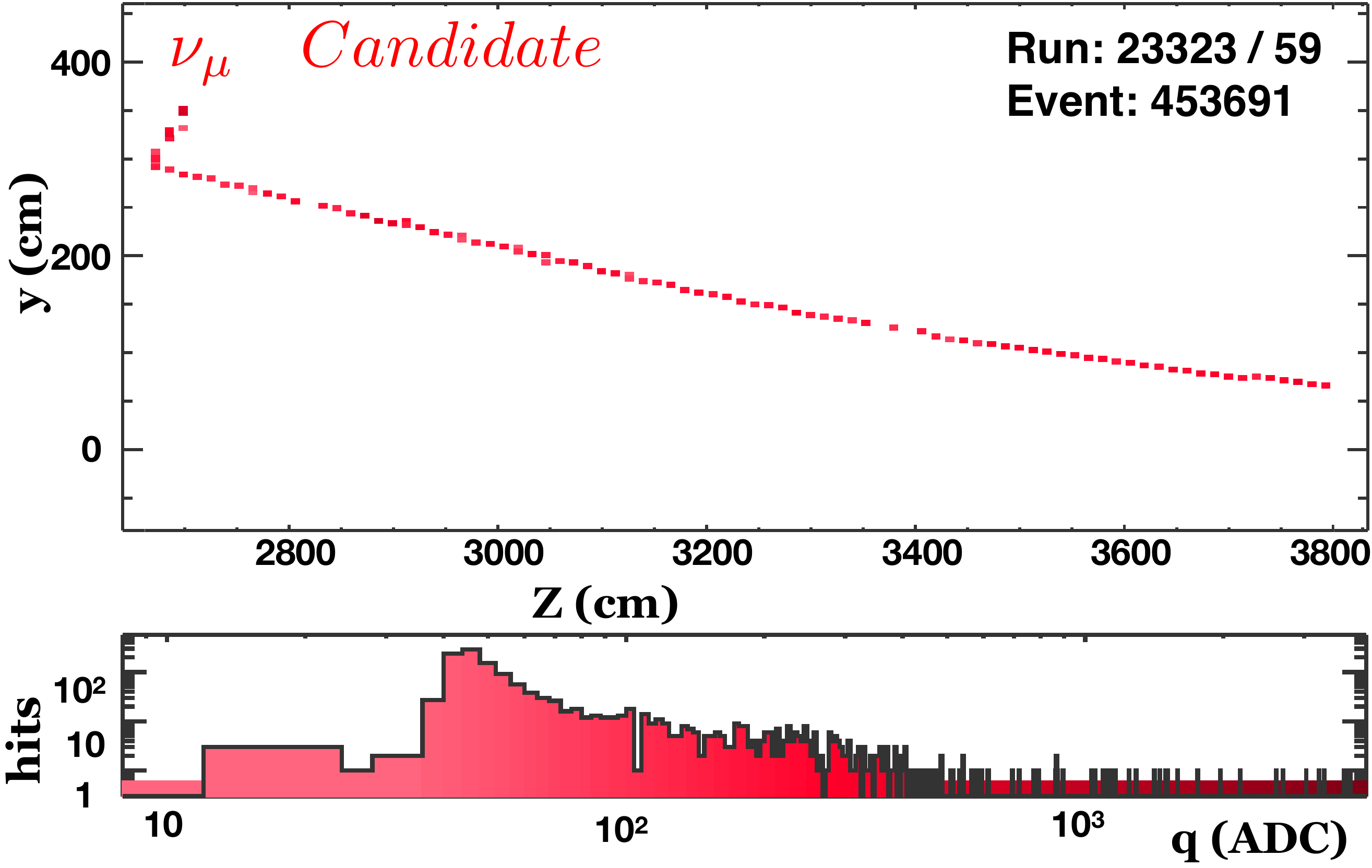}
\includegraphics[width=0.49\textwidth]{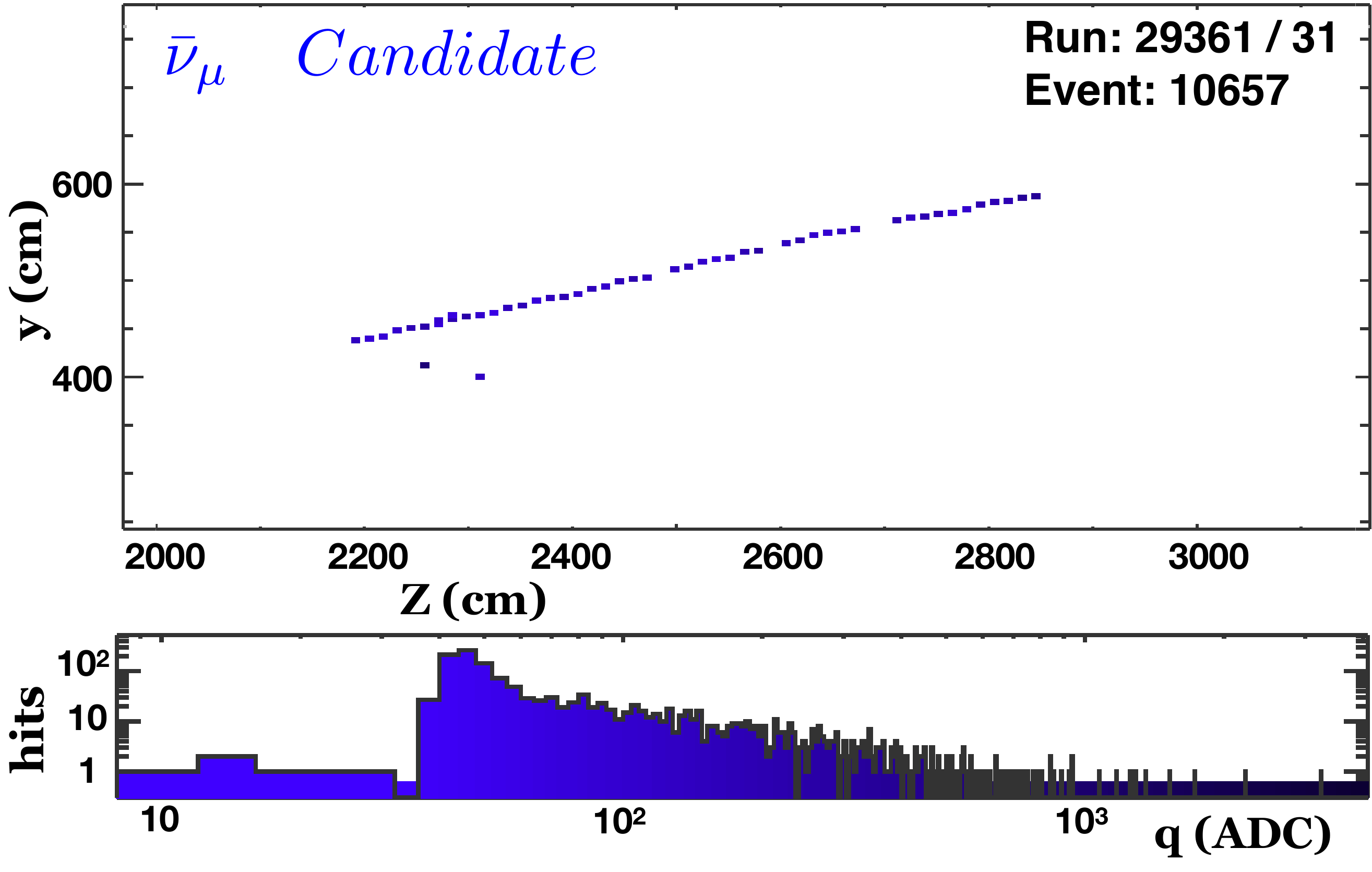}
\caption{A candidate $\nu_\mu$-CC and $\bar\nu_\mu$-CC events in the Y-Z view are shown on the left and right respectively. The charge deposited information of each event is shown underneath the event display. Both events exhibit the characteristic muon track of $\nu_\mu$-CC interactions.}
\label{fig:NumuCC-AntiNumuCC-Evds}
\end{center}
\end{figure}

An event display of $\nu_\mu$ and $\bar\nu_\mu$ candidate data events are shown on the left and right of Fig.~\ref{fig:NumuCC-AntiNumuCC-Evds}, respectively. Typically, $\nu_\mu$ events have a long, forward going track with recorded hits coming from $\mu^-$'s MIP interaction and hadronic activity coming from protons around the interaction vertex. Typical $\bar\nu_\mu$ event characteristics are a similar, long track with recorded hits coming from $\mu^+$'s MIP. Antineutrino events tend to have lower visible hadronic activity near the vertex due to interaction kinematics and the high neutron content of their final states, neutrons can soft absorb with no signs of recorded hits for its interaction. Due to the smaller hadronic energy, the $\mu^+$ will tend to be more aligned with the beam direction. The color scale shown underneath the event display remains proportional to the light seen in each cell of the detector: the light is turned into charge on the APD in order to be measured.\\

\begin{figure}[!htbp]
\begin{center}
\includegraphics[width=0.49\textwidth]{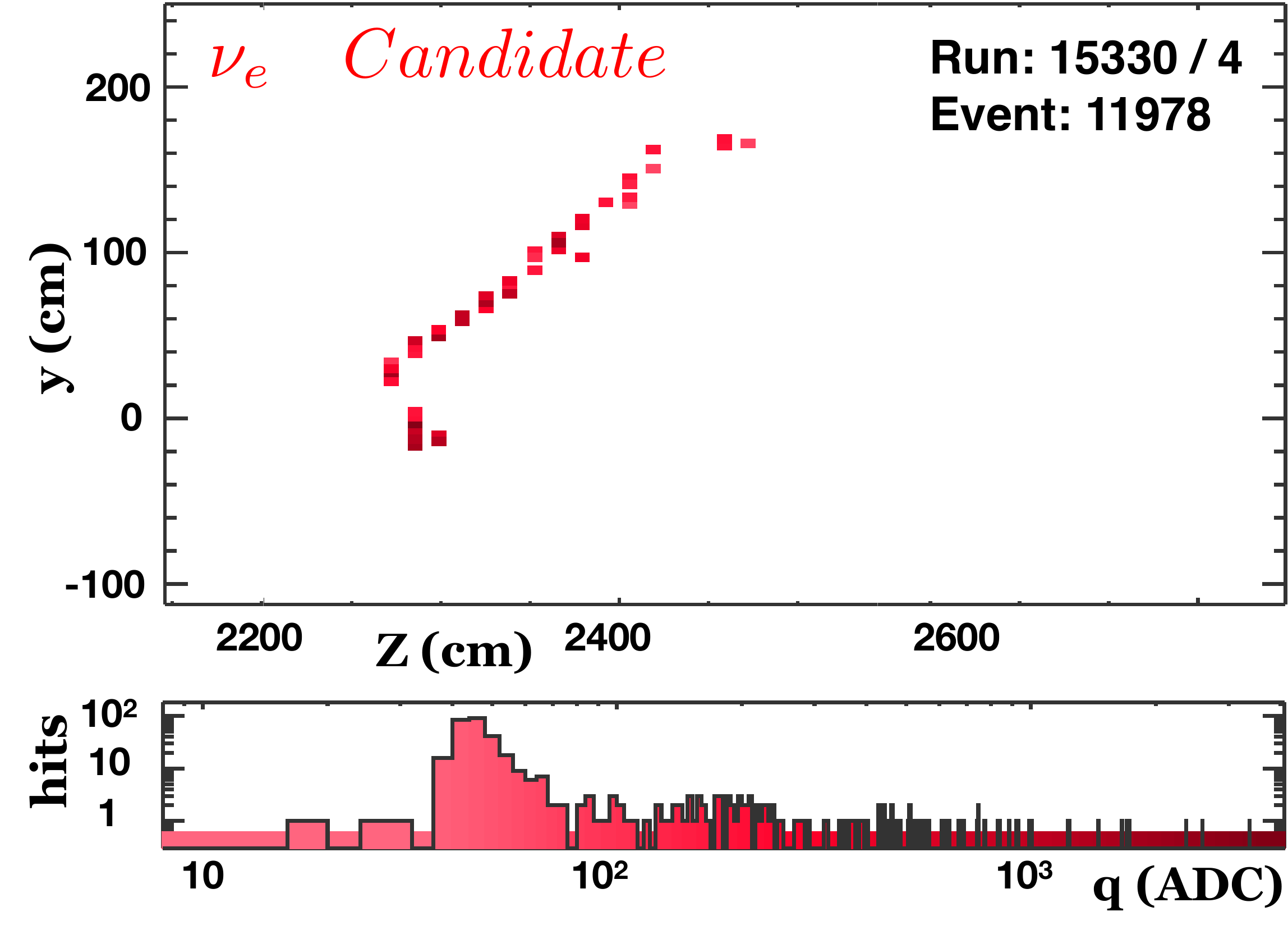}
\includegraphics[width=0.49\textwidth]{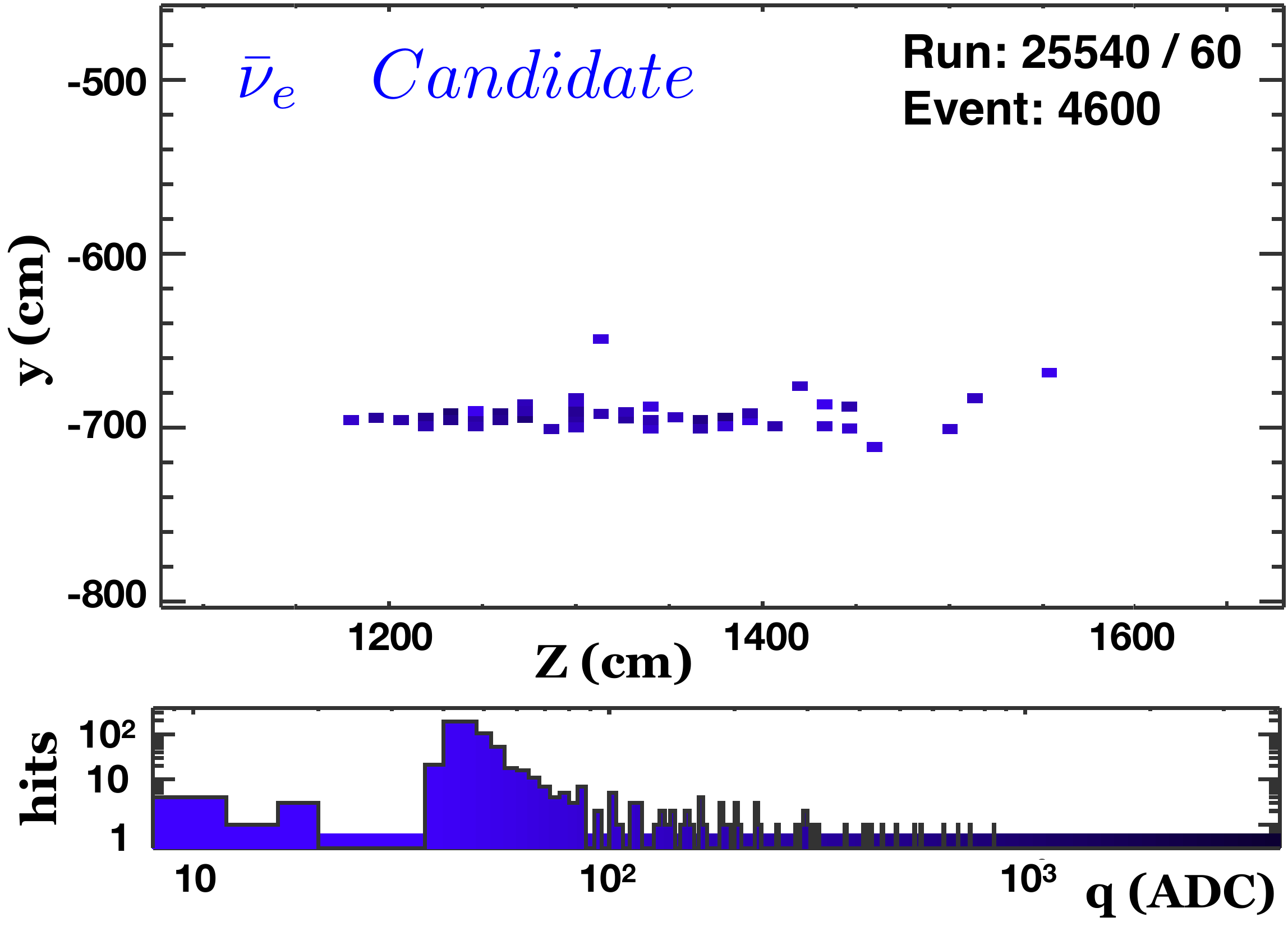}
\caption{A candidate $\nu_e$-CC and $\bar\nu_e$-CC events in the Y-Z view are shown on the left and right respectively. The charge deposited information of each event is shown underneath the event display. Both events exhibit the characteristic electron shower of $\nu_e$-CC interactions.}
\label{fig:NueCC-AntiNumuCC-Evds}
\end{center}
\end{figure}

A zoomed in event display of $\nu_e$ and $\bar\nu_e$ candidate core data events are shown on left and right of Fig.~\ref{fig:NueCC-AntiNumuCC-Evds}, respectively. Typical $\nu_e$ event characteristics are a forward going EM shower with recorded hits coming from $e^-$'s interaction and hadronic activity coming from protons around the interaction vertex, and  $\bar\nu_e$ event characteristics are an EM shower more aligned with the beam direction with recorded hits coming from $e^+$'s and, similar to $\bar\nu_\mu$, less hadronic activity around the vertex.\\

NOvA has pioneered the use of Convolutional Neural Networks (CNN) for reconstruction tasks in neutrino physics. The core of neutrino selection is the use of a CNN known as the Convolutional Visual Network (CVN), based on GoogLeNet~\cite{Aurisano:2016jvx, groh_micah_2018} a CNN for image recognition tasks. NOvA was the first experiment to utilize CNNs in a HEP result~\cite{Adamson:2017gxd}. The neutrino signal selections includes cosmic rejection, containment, data quality, and pre-selection cuts along with neutrino flavor identification from CVN.\\

Electron neutrino events passing all these selections form the "core" sample at both detectors. These events are further split into two samples of low PID and high PID score. We also construct a third, "peripheral" sample of FD events by considering events that fail containment or cosmic rejection selections, but have very high PID scores from the neural network classifier. Candidate $\nu_\mu$ events are split into four "quartiles" based on the ratio of hadronic energy to total energy in the event. \\


The neutrino energy spectrum at the NOvA ND is measured close to the neutrino source before neutrino oscillations have occurred. This large statistics data sample is used to validate the MC prediction of the expected beam flux and the simulation of the detector response. The $\nu_e$ and $\nu_\mu$ FD signal prediction is based on simulation of the $\nu_\mu$ beam flux, constrained by the observed selected $\nu_\mu$-CC candidates in the ND and oscillated appropriately.   
Discrepancies between data and MC calculations in the ND energy spectrum are extrapolated to produce a predicted FD spectrum~\cite{Suter:2015ozi,nosek_tomas}.\\

\begin{figure}[!htbp]
\begin{center}
\includegraphics[width=0.99\textwidth]{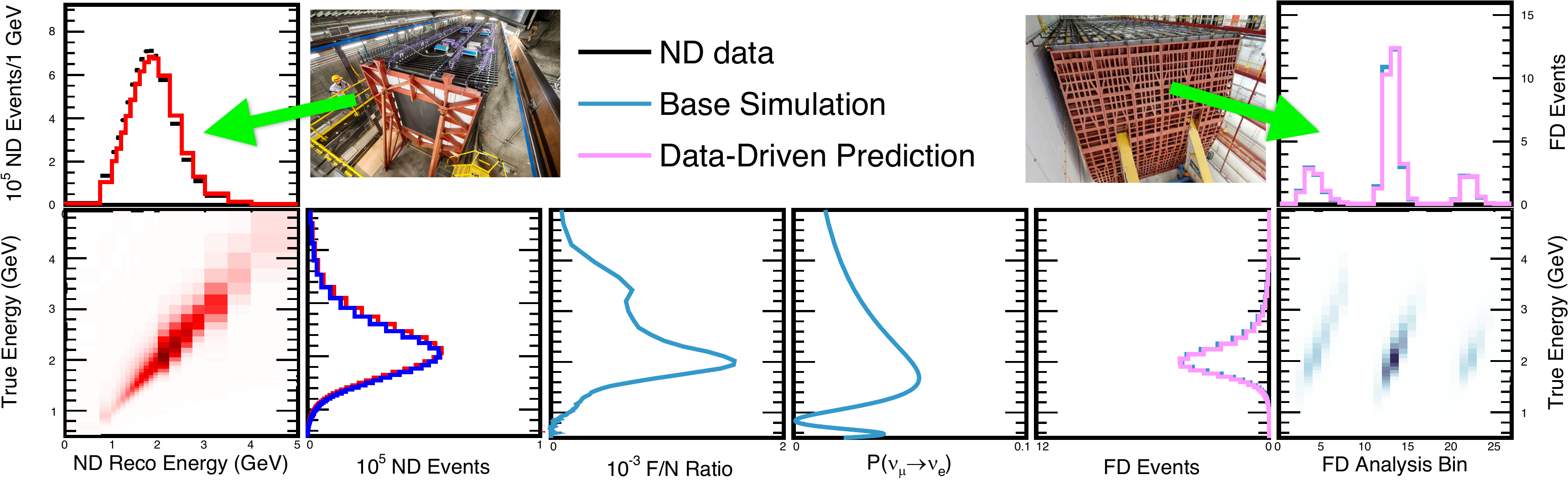}
\caption{The ND to FD extrapolation method used to predict the $\nu_{e}$-CC signal is shown here. The same extrapolation procedure is used to measure the systematic uncertainties in the oscillation analyses.}
\label{fig:extrapolation}
\end{center}
\end{figure}

The demonstration of the extrapolation procedure from the ND to the FD is shown in Fig.~\ref{fig:extrapolation}. We first convert the ND reconstructed energy spectrum into a true energy spectrum using the reconstructed-to-true migration matrix obtained from the ND simulation. The ratio of the data's unfolded spectrum to the simulated spectrum in bins of true energy is then used as a scale factor to the simulated true energy spectrum of $\nu_\mu$-CC events selected in the FD. That true energy spectrum is also weighted by the oscillation probability computed for three-flavor neutrino oscillations, including matter effects, for any particular choice of the oscillation parameters. Finally, the true energy spectrum is smeared to a reconstructed energy spectrum, again using the simulated migration matrix. In the final step, the data-based cosmic and simulation-based beam induced backgrounds are added to the prediction, which is then compared to the FD data. As the two detectors are functionally equivalent this ratio based extrapolation allows for reductions in many uncertainties, particularly beam related uncertainties and cross section uncertainties, as shown in Sec.~\ref{sec:systematics}.\\ 

\begin{figure}[!htbp]
\begin{center}
\includegraphics[width=0.99\textwidth]{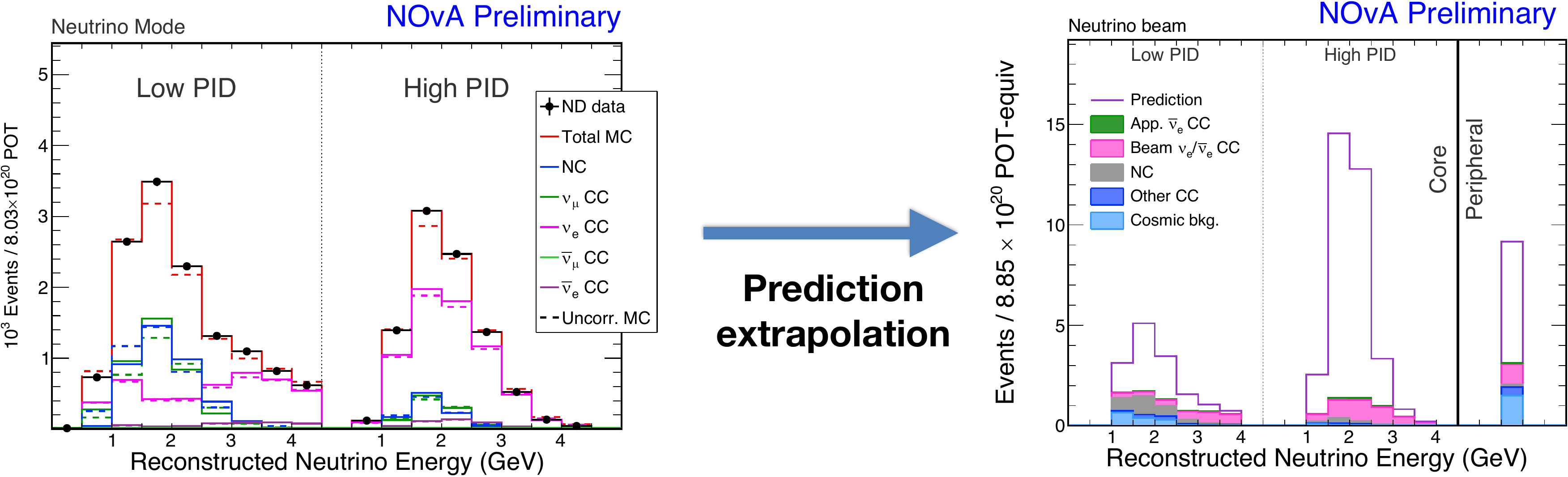}
\caption{The data-corrected near detector MC decomposed into each of the background components and the far detector prediction including the $\nu_e$-CC oscillated signal.}
\label{fig:PredictedFarDetectorSpectrum}
\end{center}
\end{figure}

The ND selected $\nu_{e}$ are oscillated to the FD by component to make a prediction of the background components. 
Each component is propagated independently in bins of energy and particle ID bins.
The cosmic background is measured using data from the NuMI timing sideband around the known beam window. A separate, 10~Hz, periodic trigger is used for zero-bias cosmic studies. The signal spectrum, background spectrum, and cosmic prediction together make up the extrapolated prediction. This is shown for neutrino mode in Fig.~\ref{fig:PredictedFarDetectorSpectrum}.\\

\section{Oscillation Systematics}
\label{sec:systematics}

We discuss the dominant systematic uncertainties associated with the joint $\nu_e+\nu_\mu$ analysis using both neutrino and antineutrino beams in NOvA. Many other effects are considered, for example the detector response modeling and normalization systematics, but will not be discussed here as the effect on event selection and final measurements is negligible. \\

\begin{figure}[!htbp]
\begin{center}
\includegraphics[width=0.99\textwidth]{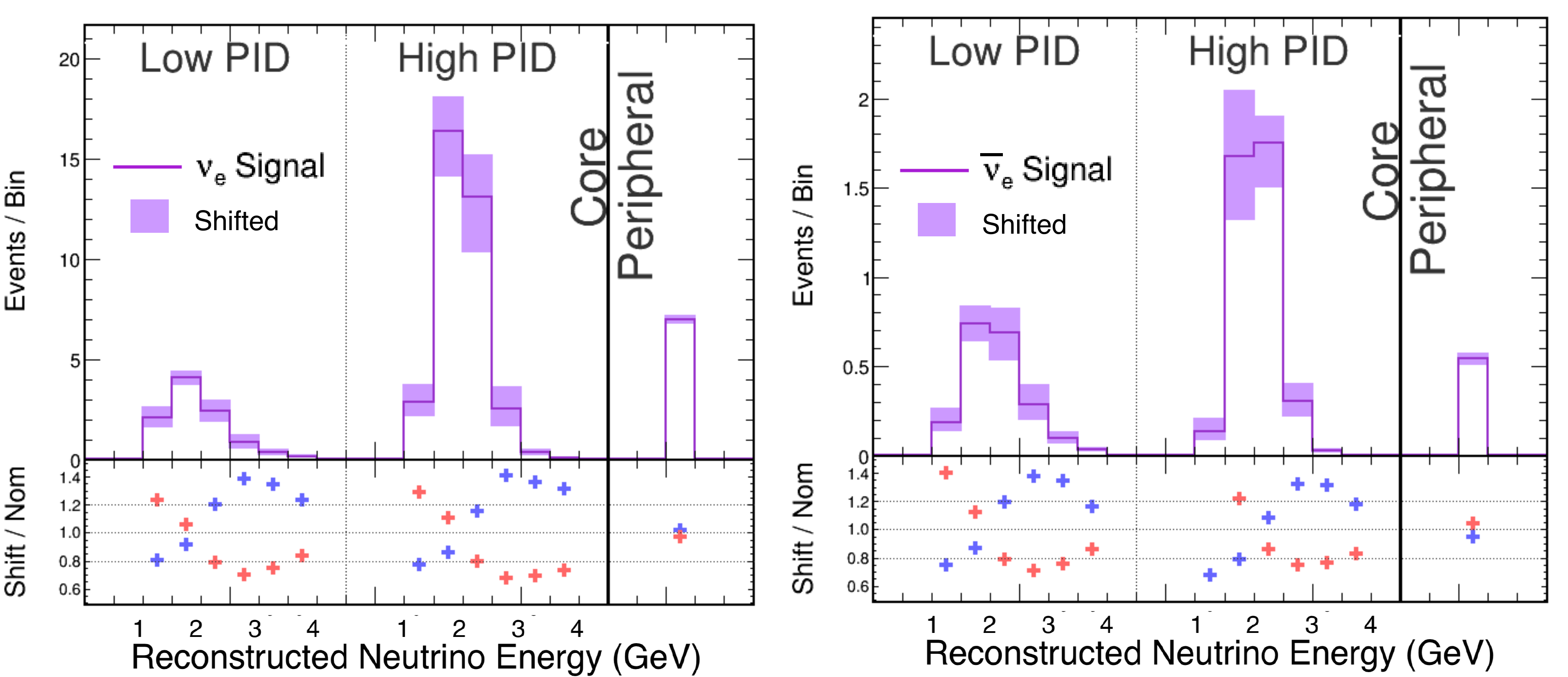}
\caption{Extrapolated FD prediction in 2 CVN bins and the peripheral for the dominant absolute calibration systematic is shown for $\nu_e$ signal (left) and $\bar\nu_e$ signal (right).}
\label{fig:Systematic}
\end{center}
\end{figure}

The impact of systematic uncertainties are estimated by producing shifted ND and FD simulation samples by event reweighing, producing specially shifted files, or altering kinematic values within events.
Systematic uncertainties in the analysis are extrapolated from the ND to the FD using the same extrapolation technique shown in Sec.~\ref{sec:oscillations}. Systematic uncertainties in the analysis are applied by replacing the systematically shifted ND spectrum in place of the nominal simulation prior to extrapolation to the FD. The systematically shifted FD predictions can then be compared to the nominal prediction for each systematic. 
As an example, the extrapolated FD prediction in 3 CVN bins for the dominant absolute calibration systematic uncertainty with the systematic error band is shown Fig.~\ref{fig:Systematic} for neutrino mode on left and antineutrino mode on right.\\

\begin{figure}[!htbp]
\begin{center}
\includegraphics[width=0.92\textwidth]{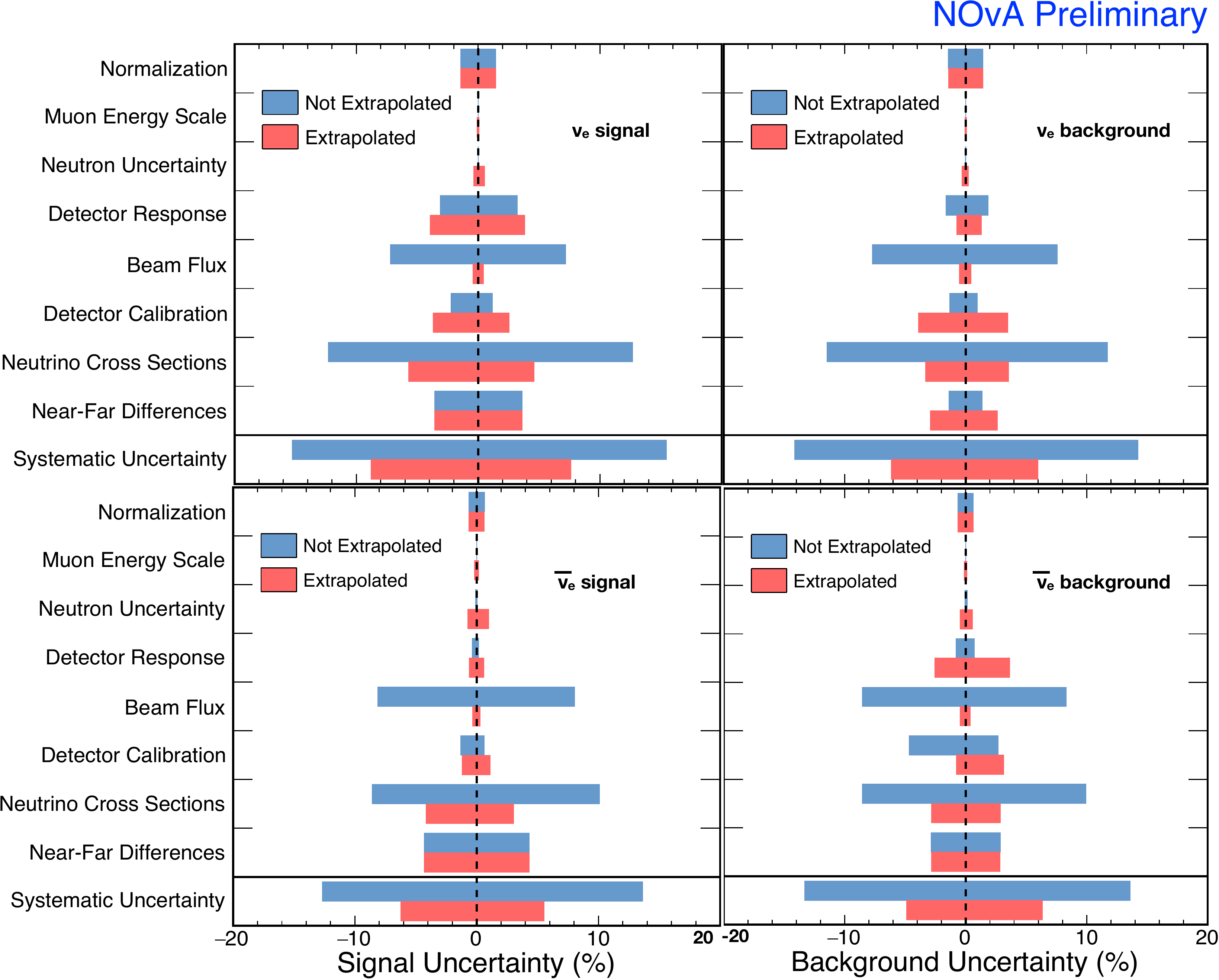}
\caption{The reduction in systematic uncertainties on the number of selected $\nu_e$ events due to using the extrapolation procedure from the ND to the FD.}
\label{fig:Syst_extrap}
\end{center}
\end{figure}

Systematic uncertainties are included as nuisance parameters in the fit in the oscillation analyses. 
In the simultaneous fit of the $\nu_e$ appearance and $\nu_\mu$ disappearance data, the nuisance parameters associated with the systematic uncertainties which are common between the two data sets, are correlated appropriately. By extrapolating from the ND to the functionally equivalent FD, the impact of many systematic uncertainties are reduced or canceled. The one sigma uncertainty in the predicted $\nu_{e}$ signal (background) events is shown in Fig.~\ref{fig:Syst_extrap}. The percentage reduction of total systematic uncertainty after the extrapolation for each component is shown in Table~\ref{tab:syst-reduction-extrap}. In particular, the variations due to uncertainties in the beam flux are almost canceled.\\


\begin{table}
\caption{The total reduction in systematic uncertainties on the number of selected $\nu_e$ events due to using the extrapolation procedure from the ND to the FD.}\label{tab:syst-reduction-extrap}
\centering%
\begin{tabular}{|l|c|c|}
\hline
 & \multicolumn{1}{|c|}{Unextrapolated} &Extrapolated \\
Component & \multicolumn{1}{c}{systematic uncertainty (\%)} & \multicolumn{1}{|c|}{systematic uncertainty (\%)} \\
\hline
$\nu_{e}$ signal&15.9&7.9\\
$\nu_{e}$ background&14.5&6.0\\
$\bar\nu_e$ signal&13.9&5.9\\
$\bar\nu_e$ background&13.9&7.0\\
\hline
\end{tabular}
\end{table}

\begin{figure}[!htbp] 
  \begin{subfigure}[b]{0.5\linewidth}
    \centering
    \includegraphics[width=0.96\linewidth]{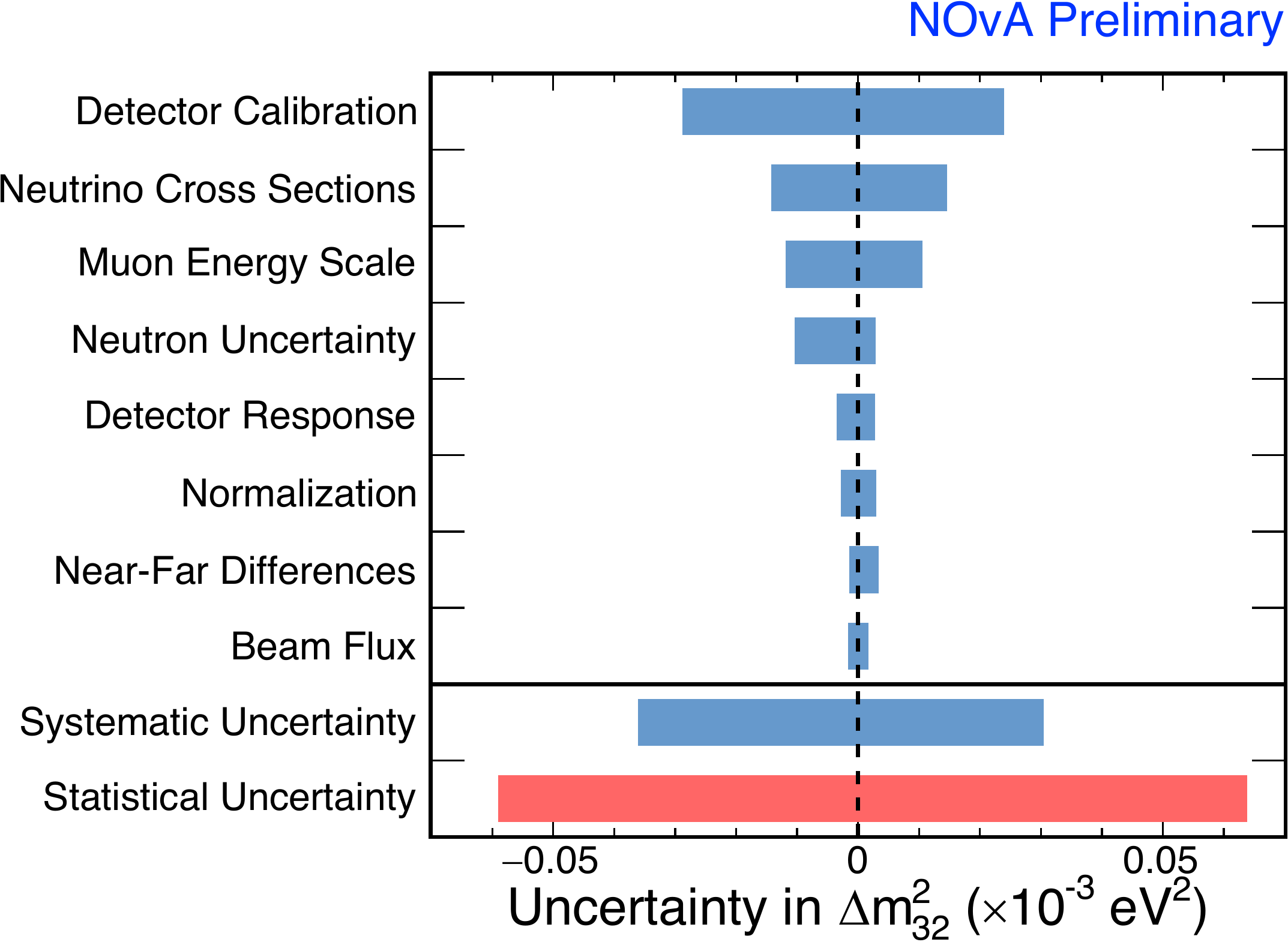} 
    \end{subfigure}
  \begin{subfigure}[b]{0.5\linewidth}
    \centering
    \includegraphics[width=0.96\linewidth]{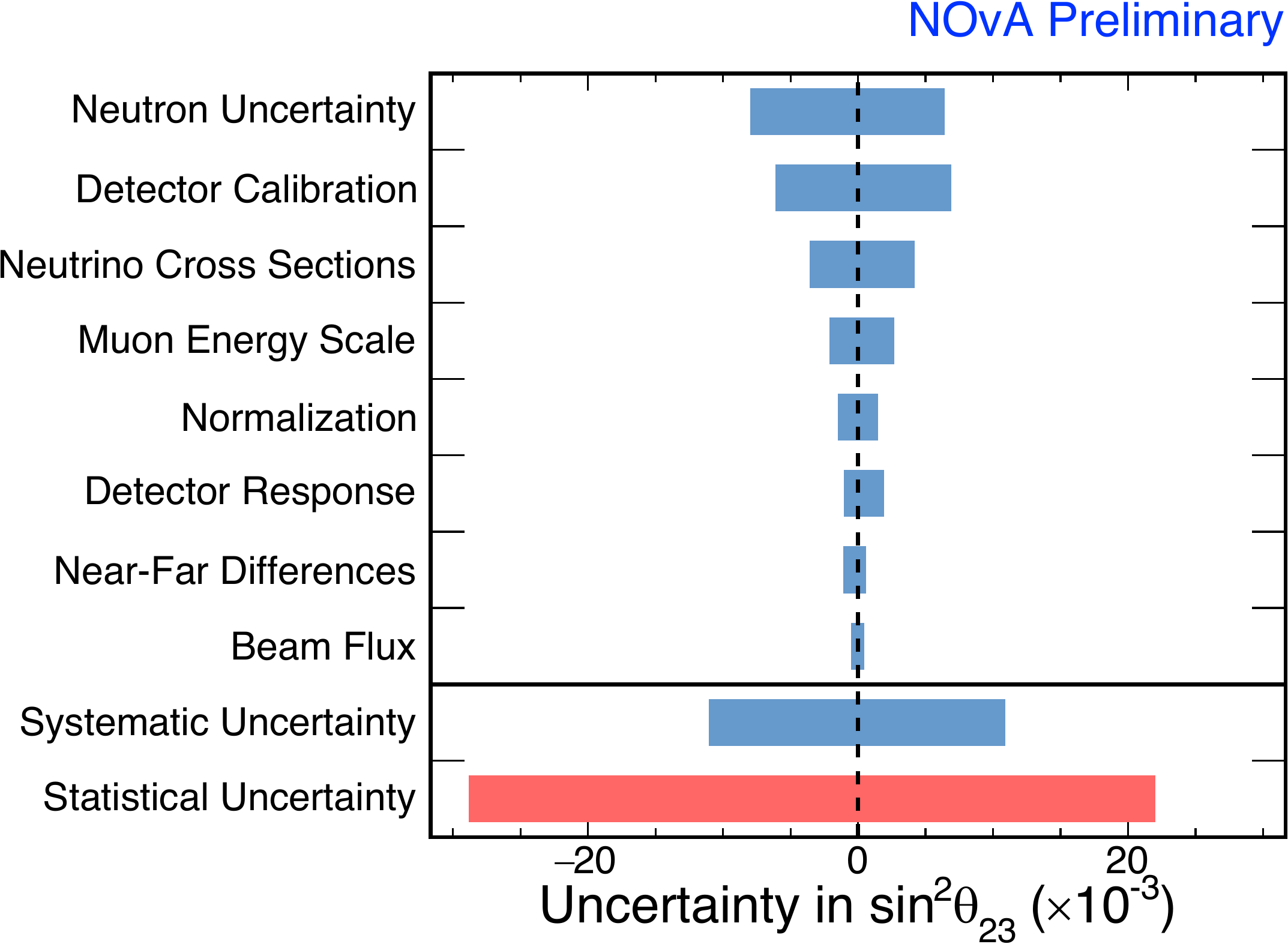} 
  
  \end{subfigure}
  \hfill\\
    \centering
\begin{subfigure}[b]{0.5\linewidth}
    \centering
  \includegraphics[width=0.96\linewidth]{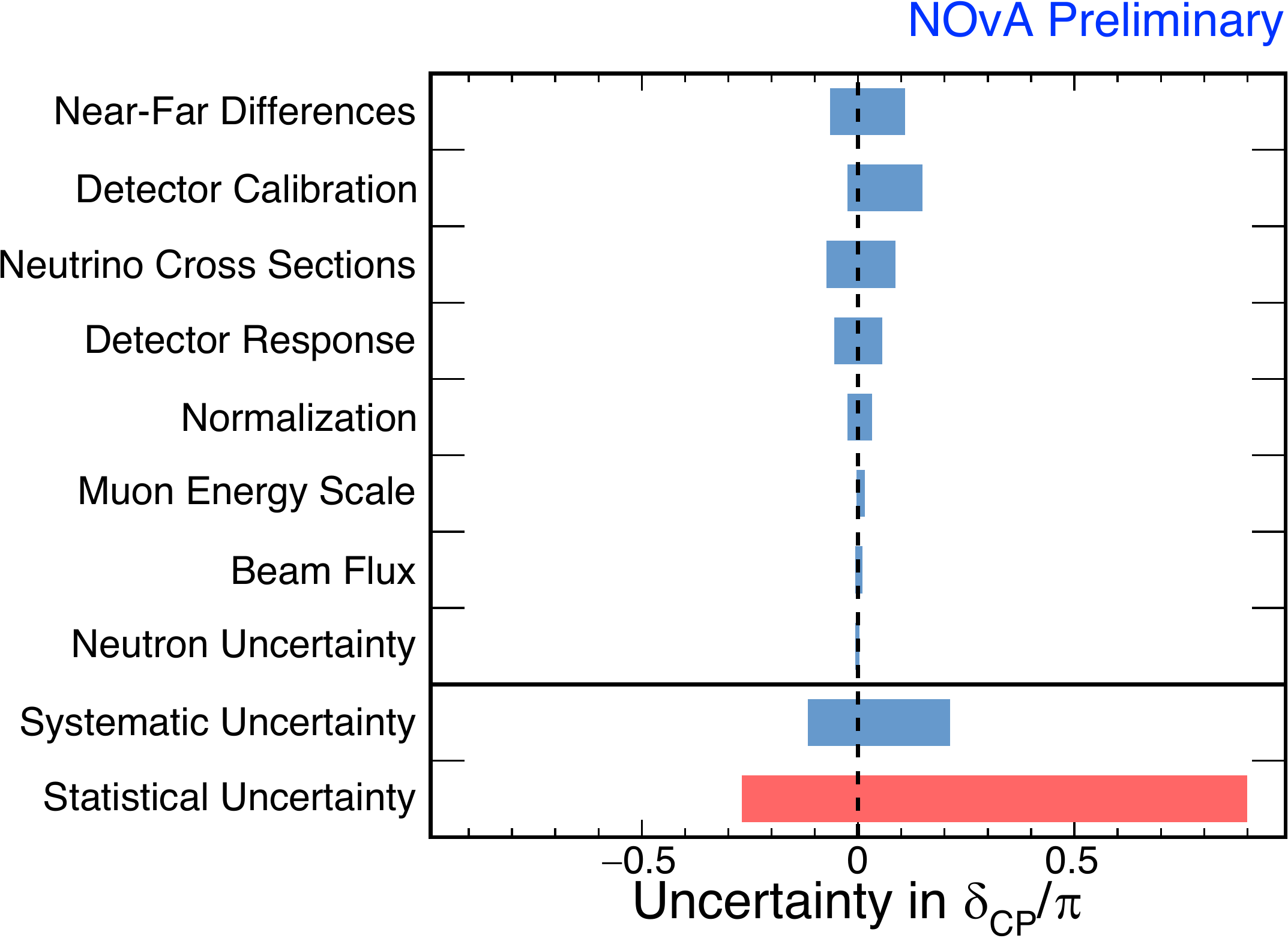} 
  \end{subfigure}
  \caption{Sources of systematic uncertainties for the oscillation parameters are shown here.}
  \label{fig:SystematicsonOsciParam} 
\end{figure}

The dominant systematic uncertainties to the joint analysis are Detector Calibration, Neutrino cross-sections, Muon energy scale, and Neutron uncertainty as shown in Fig.~\ref{fig:SystematicsonOsciParam}. These combine to contribute over 95\% of the uncertainty to $\mathrm{sin}^2\theta_{23}$ and $\Delta m^2_{32}$. The measurements are still dominated by statistical uncertainties.\\

Calibration uncertainties were assessed by the introduction of deliberate miscalibrations to the MC prior to reconstruction. These artificial miscalibrations take the form of an absolute calibration shift of all cells and a calibration shift as a function of position along cell length (separate for x and y views). The calibration uncertainties are evaluated separately for the near and far detectors. For this reason, the detector calibration and response systematics are increased when using the extrapolation method. Detector calibration and response systematics will be improved by the 2019 test beam program which aims to improve our understanding of detector calibration and response\cite{NOva-TestbeamProg}.\\


Cross section systematics are evaluated by reweighing events in the MC. Some cross section uncertainties come from the event reweighing tools built into GENIE\cite{genie}. Other uncertainties are drawn from observations made by NOvA using data from the near detector as well as external guidance from recent theory work and cross section measurements from other experiments\cite{NeutrinoXsecTune}. The cross section uncertainties considered fall into three categories: primary process (quasielastic scattering, resonance production, deep inelastic scattering, and concomitant nuclear effects like multinucleon knockout), hadronization of parton-scattering processes, and final state interactions (hadron absorption, rescattering, etc. as particles exit the nucleus). There are over 80 cross section systematics evaluated for NOvA. The largest cross section uncertainties to the $\nu_e$ and $\nu_\mu$ oscillation analyses are used directly in the oscillation fit. The remaining cross section uncertainties are used to generate principal components which are then used in the fit.\\

Principle Component Analysis (PCA) is used on NOvA to decorrelate and reduce the number of the remaining cross section systematics. The systematics are used to create an ensemble of universes in which each systematic is shifted randomly. Each universe is broken down into samples at the ND and the corrsponding F/N ratios. In practice, the uncertainties in the analysis are driven by the F/N ratios used in the extrapolation. PCA is then used to break down the RMS of the universe ensemble into principal components (PC) by diagonalizing the covariance matrix constructed from the variance within each sample. For this analysis, the largest five PCs were used with a scale factor to cover the RMS of the systematic universes.\\

PCA is also used to evaluate systematics related to operation of the beam. Two types of systematics are considered. The first account for differences between operation of the NuMI beam and the simulation of the beam. These include the horn current and position, the target position, and the beam spot size. The second type are uncertainties in the hadron production, pions and kaons, at the beam target. The NuMI beam flux is tuned using the Package to Predict the FluX using external data~\cite{Aliaga:2016oaz}. Both types of uncertainties are used to produce principal components in a similar manner to be used in the fit. \\


On NOvA, Muon energy is reconstructed using a piecewise, linear fit between the reconstructed length of the muon track and the true energy of the muon. A systematic was considered on the correspondence between muon range and energy within the NOvA detectors. Both the absolute error in each detector and the error on the ratio used in extrapolation is considered. Many sources of uncertainty were considered, but the errors are dominated by the parameterization of the density effect and the mass accounting in the detector.\\

\begin{figure}[!htbp]
\begin{center}
\includegraphics[width=0.96\textwidth]{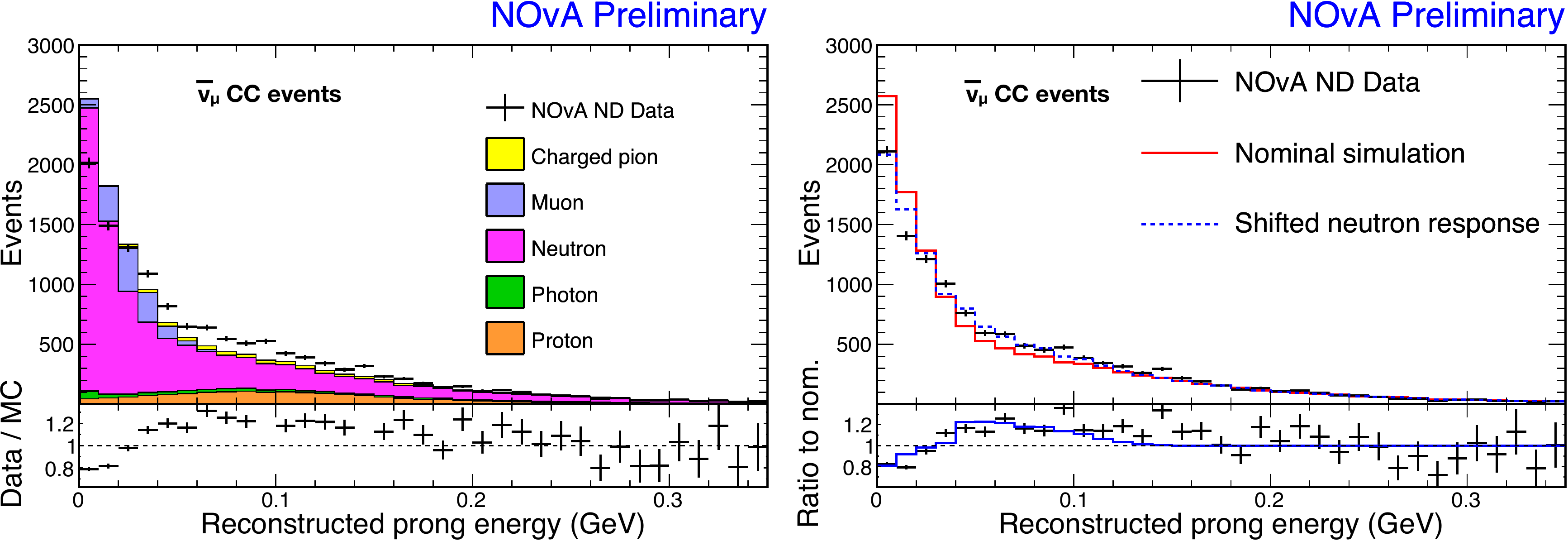}
\caption{Prongs are broken down in selected $\bar\nu_\mu$ CC events by the particle, or parent if the parent was a neutrino, that deposited energy in the event. The selection used here is to select a neutron rich sample.}
\label{fig:NeutronSystematic}
\end{center}
\end{figure}

An uncertainty in the response of the detector to neutrons is new in the antineutrino oscillations analysis. Antineutrino charge current interactions are much more likely to produce final-state neutrons often with several hundred MeV of energy, while neutrino charged current interactions produce final state protons. 
Modeling these fast neutrons is known to be challenging due to the lack of tagged neutron data. A highly selected neutron rich $\bar\nu_\mu$ CC Quasi-elastic like event sample with neutron angle consistent with the QE hypothesis is selected from NOvA ND data and MC.  There is a discrepancy in the total calorimetric energy of reconstructed neutron prongs from $\bar\nu_\mu$ CC events shown in the left of Fig.~\ref{fig:NeutronSystematic}. A new systematic is introduced which scales the amount of deposited energy of some neutrons to cover the low-energy discrepancy. This scaling shifts the mean $\bar\nu_\mu$ energy by 1\% and the mean $\nu_\mu$ energy by 0.5\%~\cite{tyler_numusyst}.

\section{Cross-Checks with Muon-Removed Electron-Added Sample}

\label{sec:mre}
The low number of $\nu_{e}$ events makes the cross-checks of the modeling of the hadronic component in the $\nu_e$ signal challenging. Muon-Removed Electron-Added (MRE) is a unique data-driven technique for this channel that uses the large statistics of the $\nu_{\mu}$ CC data events in the ND data sample. A muon-removal algorithm~\cite{Sachdev:2013ema} replaces the muon in selected $\nu_{\mu}$ CC events with a simulated electron of the same energy in Data and Monte Carlo while preserving the nuclear/hadronic portion of the interaction.  These hybrid Data/Monte Carlo events allows us to study the impact of any mis-modeling of the hadronic shower on the $\nu_{e}$ selection efficiency.\\

\begin{figure}[!htbp]
\begin{center}
\includegraphics[width=0.99\textwidth]{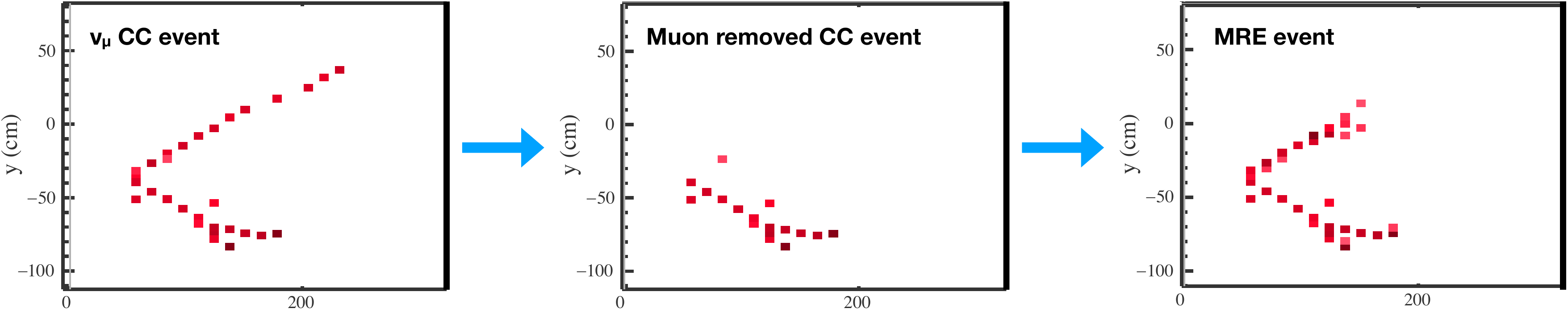}
\caption{Muon-removed electron-added technique. The ND candidate data $\nu_\mu$-CC event is shown at the left, the muon-removed CC event is shown in middle, and the muon-removed simulated electron (with same energy and direction) replaced event display is shown at the right. }
\label{fig:MRE}
\end{center}
\end{figure}

\noindent MRE event generation has three steps as shown in Fig~\ref{fig:MRE}: 
\begin{enumerate}[label=\textbf{\arabic*}.]
\item{\textbf{Generating a muon-removed charged-current (MRCC) sample:} We select the muon track in the event using the muon PID  and remove all its associated track hits. Far from the vertex, muon track hits are clean, however, for hits close to the vertex, given that there may be some contamination from the hadronic energy, only a minimum ionizing particle energy (MIP) from each hit near the vertex is removed.}
\item{\textbf{Generation of the electron:} Once the muon is removed from the event, an electron with the same starting point, direction, and reconstructed energy of the original muon track is simulated in its place using the standard NOvA GEANT4 tool.}
\item{\textbf{Creating a muon-removed, electron-added event:} The simulated electron hits are overlaid with the hits of the MRCC event and the NOvA reconstruction algorithm runs from the beginning.}
\end{enumerate}

\begin{figure}[!htbp]
\begin{center}
\includegraphics[width=0.49\textwidth]{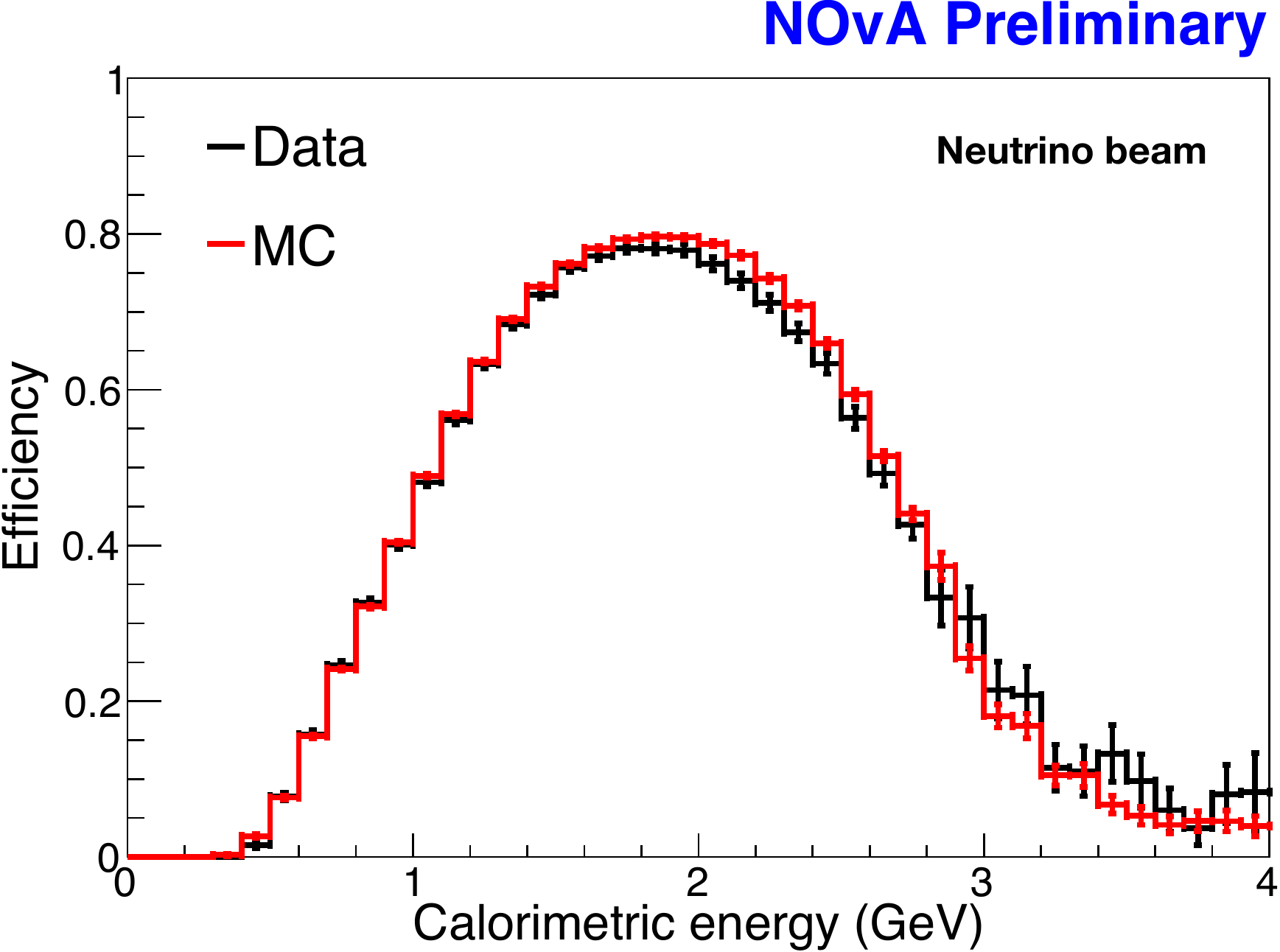}
\includegraphics[width=0.49\textwidth]{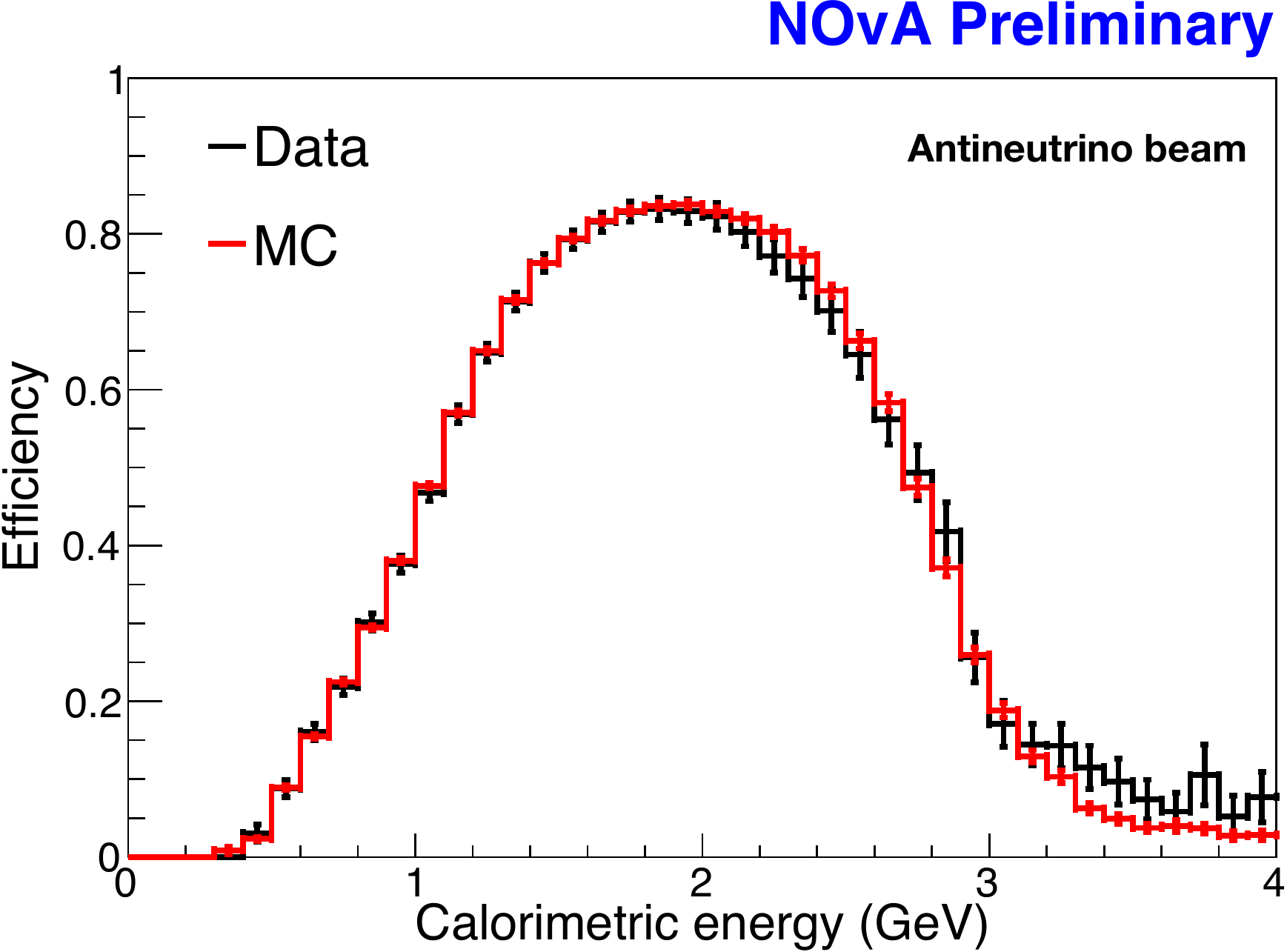}
\caption{The PID selection efficiency as a function of calorimetric energy is shown.}
\label{fig:MRECalE}
\end{center}
\end{figure}

The CVN $\nu_e$ and $\bar\nu_e$ events selection efficiencies in Data and Monte Carlo with respect to the pre-selection are compared in Fig~\ref{fig:MRECalE} using the  CVN $\nu_e$ selection efficiency on the calorimetric energy (at the left for $\nu_e$ and at the right for $\bar\nu_e$). The data and MC total efficiencies agree at the $2\%$ level for MRE events both in neutrino and antineutrino beams. 

\section{Cross-Check using Muon Removed Bremsstrahlung Showers}
\label{sec:mrbrem}
\begin{figure}[!htbp]
\begin{center}
\includegraphics[width=0.97\textwidth]{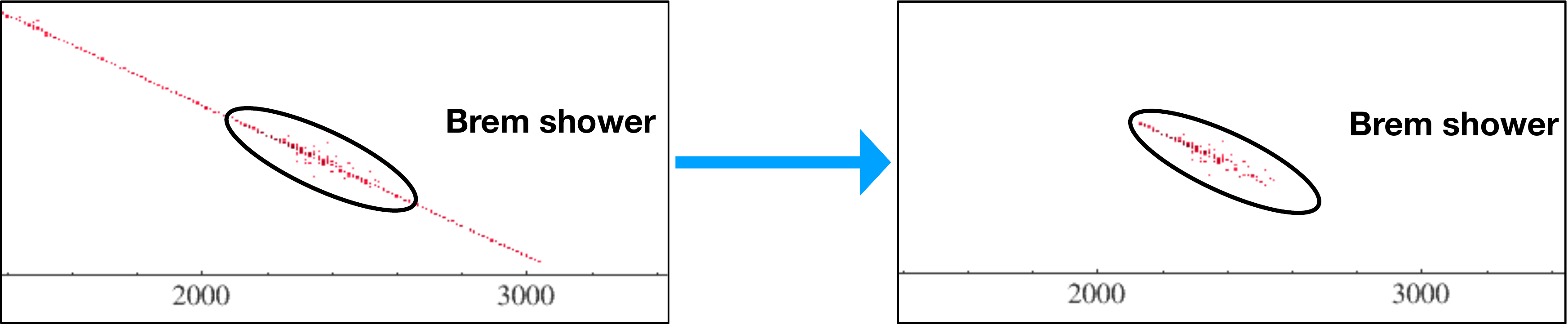}
\caption{Bremsstrahlung shower from cosmic moun before (left) and after (right) muon removal.}
\label{fig:MRBrem}
\end{center}
\end{figure}

The $\nu_e$ ($\bar\nu_e$) events are identified by the electromagnetic showers induced by $e^-$, or $e^+$, in the final state interactions. The NOvA FD sits on the surface, but it is covered by 3~m of barite rock and concrete which provides an overburden of more than ten radiation lengths to reduce background from cosmic rays. Still, there is a high rate (148 kHz) of cosmic muons observed in the FD. These muons can induce EM showers by three different means: energetic muons undergoing bremsstrahlung radiation (Brem), muons decaying into electrons in flight (DiF), and muons stopping in the detectors and decaying into Michel electrons. Michel electrons typically have energies much smaller than $\nu_{e}$ events, which have energies of 0.5~GeV \-- 4~GeV, and have instead been used as a calibration check. Brem and DiF, on the other hand, provide abundant EM showers in the few-GeV energy region. A pure sample of EM showers is obtained from cosmic data through a modified EM shower filtering algorithm~\cite{Duyang:2015cvk} from the Muon-Removal (MR) algorithm~\cite{Sachdev:2013ema} for charged current events. This sample can be used to characterize the EM signature and provide valuable cross-checks of the MC simulation, reconstruction, CVN algorithms, and calibration of the FD~\cite{Duyang:2015cvk}. The results shown in this section are from Brem showers only. DiF events requires additional event selections which gives lower statistics, but more pure EM samples, and are not included in this section. 
\\\\
The EM shower filtering algorithm first looks at events for a cosmic EM shower, which has a muon track inside the EM shower region. Then, the MR algorithm removes hits that belong to the muon corresponding to the energy of a MIP in the shower region and saves the remaining EM showers as raw DAQ hits. An example EM shower event display before and after MR from FD cosmic data is shown in Fig.~\ref{fig:MRBrem}. The shower digits can then be put into standard $\nu_e$ reconstruction and PID algorithms. Data and MC comparison is performed with reconstructed shower variables and PID outputs to validate EM shower modeling and PID. Calibration effects can be checked by comparing PID efficiencies as a function of vertex position.
\\

\begin{figure}[!htbp]
\begin{center}
\includegraphics[width=0.96\textwidth]{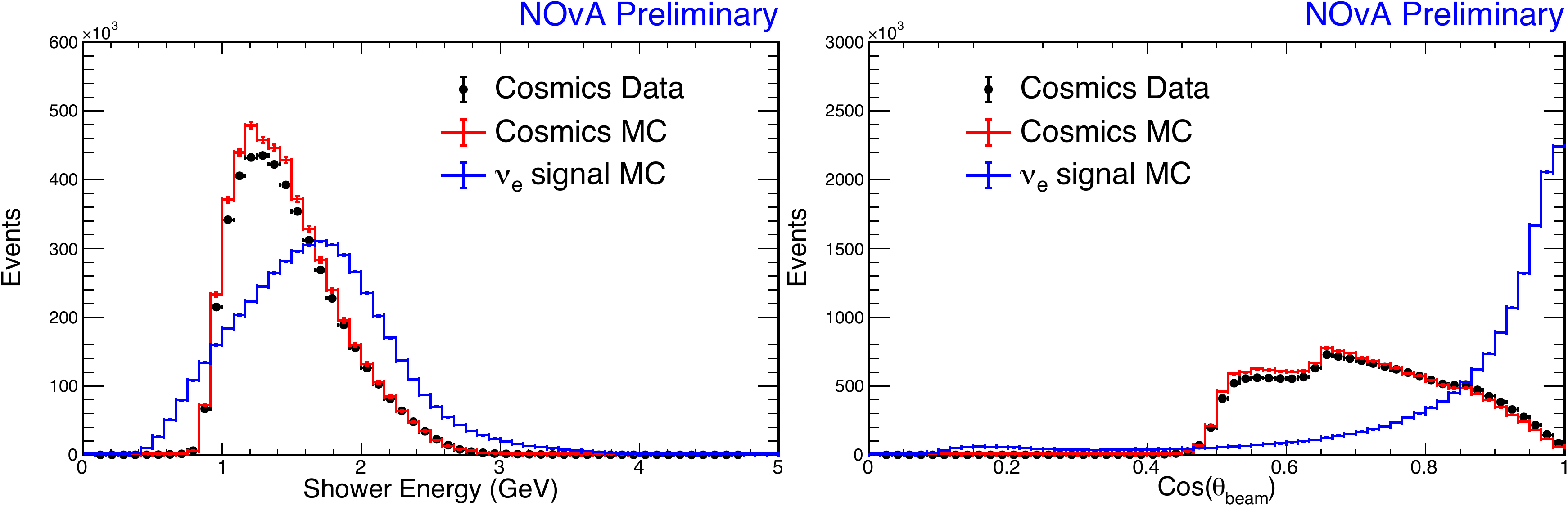}
\caption{Brem shower energy vs $\nu_e$ shower energy is shown on left, and the comparison of shower angle $\mathrm{cos(\theta_{beam})}$ is shown on right.}
\label{fig:MRBremEnergyangle}
\end{center}
\end{figure}

\begin{figure}[!htbp]
\begin{center}
\includegraphics[width=0.47\textwidth]{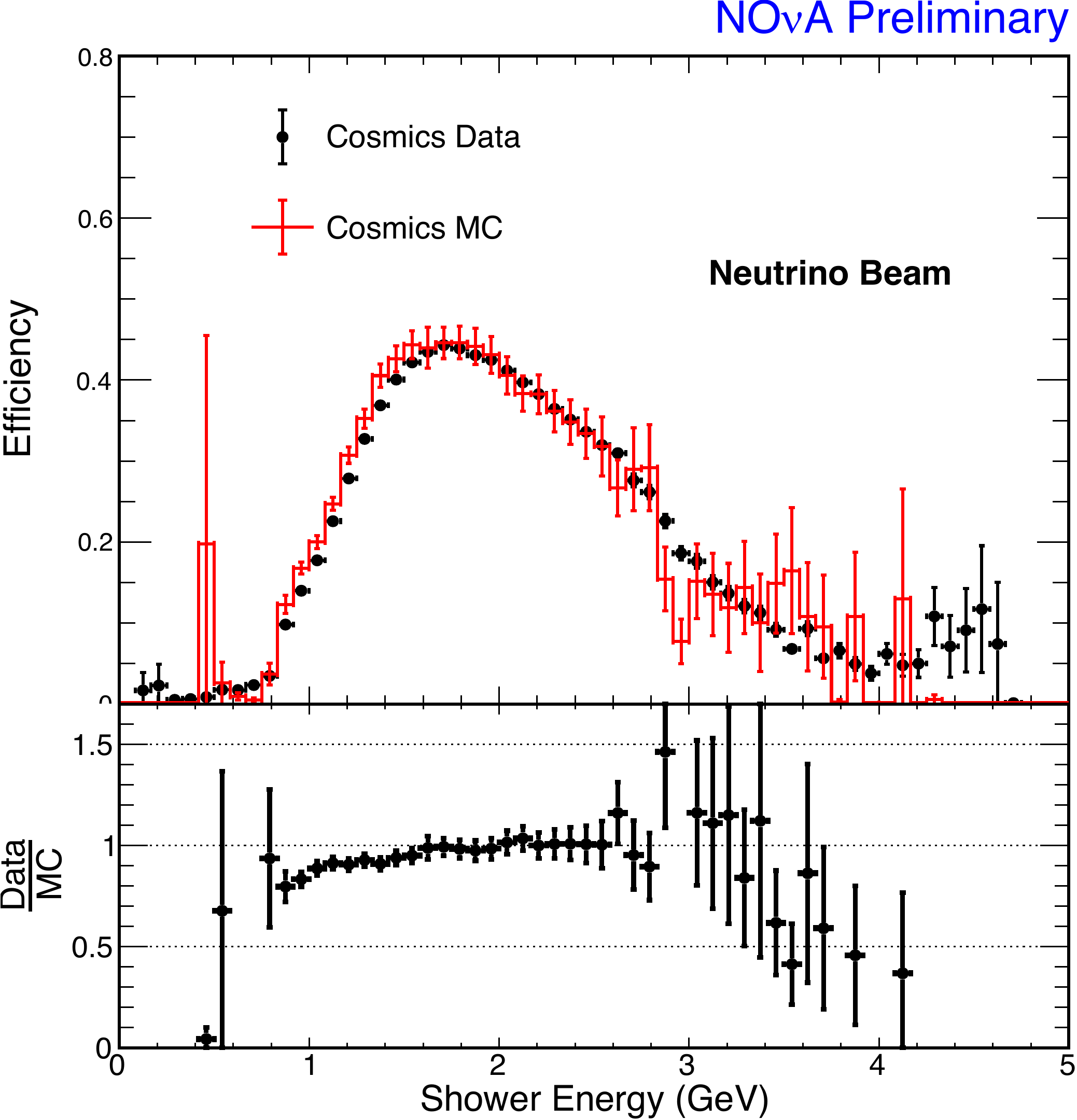}
\includegraphics[width=0.52\textwidth]{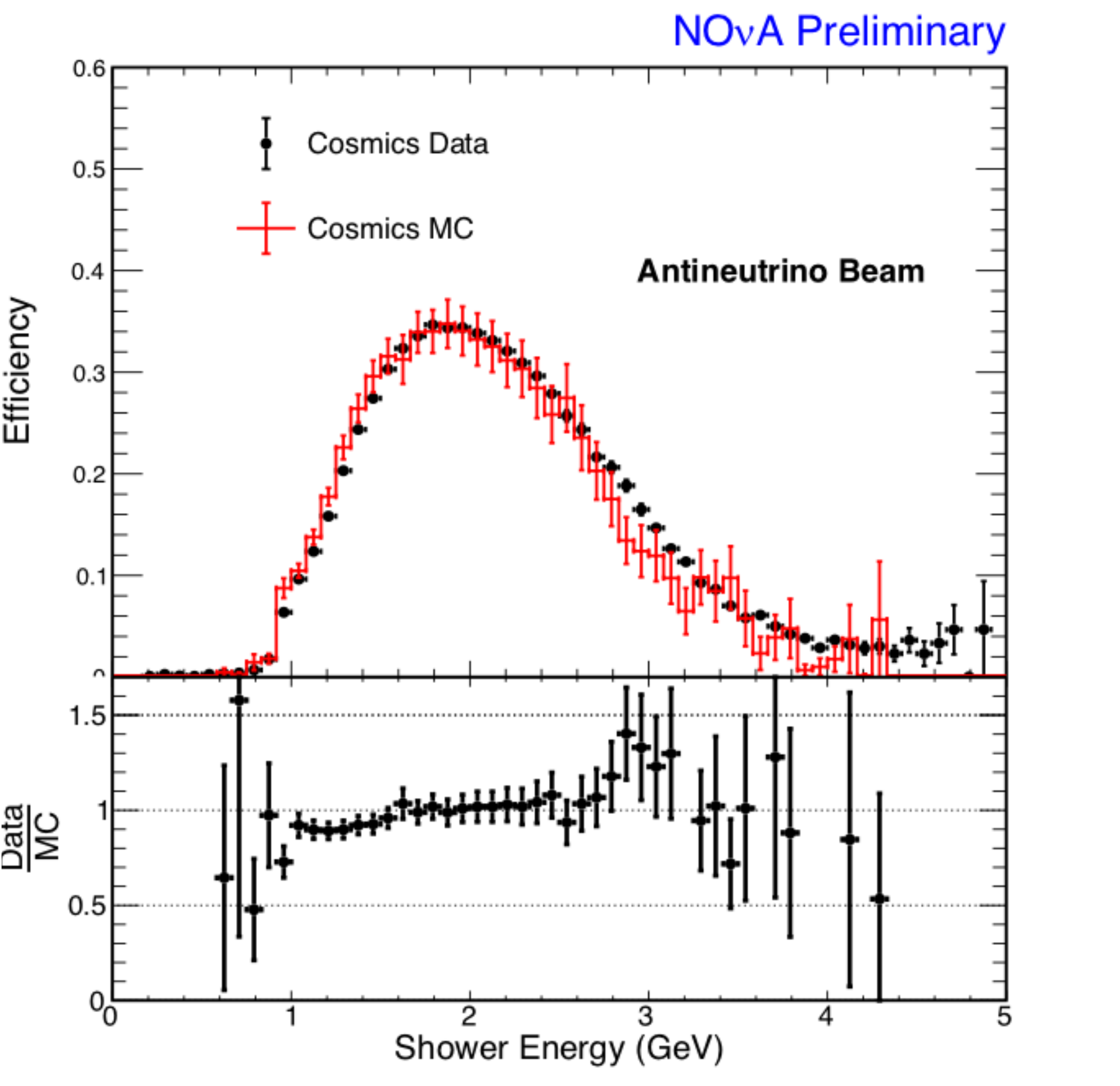}
\caption{The PID selection efficiency as a function of EM showers energy is shown.}
\label{fig:MRBremShowerE}
\end{center}
\end{figure}

\noindent MR Brem Shower extraction takes place through the following steps:
\begin{enumerate}[label=\textbf{\arabic*}.]
\item{\textbf{Muon track selection:} Apply selections to search for a cosmic muon candidate in the FD. A muon track should be long enough to generate bremsstrahlung showers, thus require the number of planes that the muon track traverses to be greater than 30. The muon track is also required to be in the horizontal direction requiring $\mathrm{cos\theta}>0.5$, where $\mathrm\theta$ is the angle of the muon track with respect to the beam axis (Z-axis).}
\item{ \textbf{Shower finding:} The shower region is found within the muon track candidate. The shower region is determined by measuring the energy deposition per plane (dE/dx) information.}
\item {\textbf{Muon removal:} Once the EM showers are identified,
muon hits are removed to get EM shower events.}
\item{\textbf{Shower Reconstruction and PID:} The EM shower events are fed into standard $\nu_e$ reconstruction and PID algorithms.}
\end{enumerate}

\noindent Reconstructed shower energy and angle comparison for Brem showers and $\nu_e$ signal induced  EM showers are shown on top plots of Fig.~\ref{fig:MRBremEnergyangle}. The main differences between $\nu_e$ signal and Brem sample is that beam-related $\nu_e$ energy peaks at 2 GeV and its direction is along the NuMI beam line direction whereas cosmic EM shower is mostly coming from vertically into the FD. As a result the CVN selection efficiency is low for muon removed brem showers. To correct this, we reweigh the Brem sample according to $\nu_e$ signal angle to resemble the sample.\\

The CVN $\nu_e$ and $\bar\nu_e$ event selection efficiencies with respect to the pre-selection are calculated to compare Brem data and cosmic simulation. The CVN selection efficiency of EM shower energy in $\nu_e$ and $\bar\nu_e$ events is shown in left and right of Fig~\ref{fig:MRBremShowerE}, respectively. The Data and MC CVN selection efficiency in core sample agrees well within 6\% (3\%) level in neutrino (antineutrino) mode. 
Efficiency of data and simulated Brem showers agrees within the total extrapolated systematic uncertainties shown in Fig.~\ref{fig:Syst_extrap} for neutrino and antineutrino datasets.
\section{Summary and Conclusions}
\label{sec:summary}
 NOvA has analyzed it's first $6.9\times10^{20}$ POT antineutrino data together with that of it's $8.85\times10^{20}$ POT neutrino data to set new constraints on neutrino oscillation parameters, and observed $>4\sigma$ evidence of  $\bar\nu_e$ appearance. The study of systematic uncertainties and muon-removed cross-checks are crucial elements of this analysis. 
The efficiency agrees between data and MC at the $2\%$ level for MRE events both in neutrino and antineutrino beam modes and the efficiency of data and simulated MRBrem showers agrees within systematics for neutrino and antineutrino datasets.                                                           
 
\section{Acknowledgments}
This work was supported by the US Department of Energy; the US National Science Foundation; the Department of Science and Technology, India; the European Research Council; the MSMT CR, Czech Republic; the RAS, RMES, and RFBR, Russia; CNPq and FAPEG, Brazil; and the State and University of Minnesota. We are grateful for the contributions of the staffs at the University of Minnesota module assembly facility and Ash River Laboratory, Argonne National Laboratory, and Fermilab. Fermilab is operated by Fermi Research Alliance, LLC under Contract No. De-AC02-07CH11359 with the US DOE.

\bigskip 

\end{document}